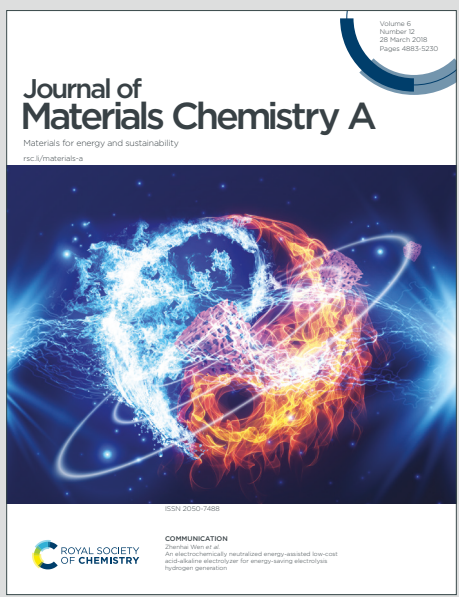

# Combinatorial screening of crystal structure in Ba-Sr-Mn-Ce perovskite oxides with $ABO_3$ stoichiometry

Su Jeong Heo*[a] and Andriy Zakutayev*[a]

[a]National Renewable Energy Laboratory, Golden, CO 80401, USA

**Corresponding Authors**

*Su Jeong Heo: SuJeong.Heo@nrel.gov

*Andriy Zakutayev: Andriy.Zakutayev@nrel.gov

**Abstract**

ABO$_3$ oxides with the perovskite-related structures are attracting significant interest due to their promising physical and chemical properties for many applications requiring tunable chemistry, including fuel cells, catalysis, and electrochemical water splitting. Here we report on the crystal structure of the entire family of perovskite oxides with ABO$_3$ stoichiometry, where A and B are Ba, Sr, Mn, Ce. Given the vast size of this chemically complex material system, exploration for stable perovskite-related structures with respect to its constituent elements and annealing temperature is performed by combinatorial pulsed laser deposition and spatially-resolved characterization of composition and structure. As a result of this high-throughput experimental study, we identify hexagonal perovskite-related polytypic transformation as a function of composition in the Ba$_{1-x}$Sr$_x$MnO$_3$ oxides after annealing at different temperatures. Furthermore, a hexagonal perovskite-related polytype is observed in a narrow composition-temperature range of the BaCe$_x$Mn$_{1-x}$O$_3$ oxides. In contrast, a tetragonally-distorted perovskite is observed across a wider range of compositions and annealing temperatures in the Sr$_{1-x}$Ce$_x$MnO$_3$ oxides. This structure stability is further enhanced along the BaCe$_x$Mn$_{1-x}$O$_3$ - Sr$_{1-x}$Ce$_x$MnO$_3$ pseudo-binary tie-line at x=0.25 by increasing Ba-incorporation and annealing temperature. These results indicate that the BaCe$_x$Mn$_{1-x}$O$_3$ - Sr$_{1-x}$Ce$_x$MnO$_3$ pseudo-binary oxide alloys (solid solutions) with tetragonal perovskite structure and broad composition-temperature range of stability are promising candidates for thermochemical water splitting applications.

## 1. Introduction

Oxides materials with general formula $ABO_3$ and perovskite-related crystal structures are widely used in various energy applications including chemical and electrochemical catalysis, photocatalysis, and ferroelectricity due to their excellent physical properties such as chemical stability, structural diversity, and electrical conductivity.[1-4] The A-site cations in the perovskites are typically alkaline earth or rare earth metals, which occupy the corners of the unit cell with a 12-fold coordination by the anions. The B-site cations are generally transition metals that are in the center of the unit cell with 6-fold coordination by the oxygen anions. The oxygen anions form an octahedral network that share corners in an ideal cubic perovskite, and the ability of the octahedra to be tilted leads to various phase transformation and corresponding changes in properties. The numerous possible substitutions on A/B cation sites with elements of different oxidation states lead to highly tunable chemical properties, structural diversity, oxygen vacancies, redox active sites, as well as electronic and ionic conductivity.[5-7]

Of particular interest to this paper, the perovskite oxides have recently emerged as redox materials for a solar-driven thermochemical hydrogen production (STCH) due to their promising properties for this application, including high oxygen exchange capacities, high structural tolerance during the cycling, and good chemical stability at relatively low operation temperatures.[8,9] The perovskites are promising STCH materials because of their better oxygen storage capacity and hydrogen fuel production at lower temperatures compared to ceria ($CeO_2$) and doped ceria (such as $Ce_{1-x}MO_{2-\delta}$).[10,11] The redox properties of the perovskite oxides ($ABO_3$) are tunable by adjusting the A- and/or B-site substitution level (also referred to as 'doping'), and demonstration of various compounds is achieved with higher fuel production than ceria for

$BaCe_{0.25}Mn_{0.75}O_{3-\delta}$,[12] $Sr_{1.8}Ce_{0.2}MnO_4$,[13] $SrTi_{0.5}Mn_{0.5}O_{3-\delta}$,[14] $LaGa_{0.4}Co_{0.6}O_{3-\delta}$,[15] $La_{1-x}Sr_xMnO_3$[16] and $Sr_xLa_{1-x}Mn_yAl_{1-y}O_{3-\delta}$.[17]

Given the large number of possible compositions for exploration within the perovskite structure, discovery of new stable perovskite structures with respect to its constituent elements and operating temperature is of great important for identifying the promising STCH material candidates. To address this challenge, high-throughput density functional theory (DFT) studies have been introduced to search for novel perovskites for STCH applications,[18] screening 5,329 cubic and distorted perovskite $ABO_3$ compounds for their thermodynamic stability and oxygen vacancy formation energy. Similarly, high-throughput experimental approach combining combinatorial synthesis and spatially-resolved characterization can be used to screen the complex perovskite oxide materials with extended variation in the compositional space. In the past, combinatorial thin films libraries with a large compositional variation demonstrated the efficiency of such high-throughput experimental approach for studying halide perovskite solar cell,[19] transition metal dopants in oxide perovskite ferroelectrics,[20] and their piezoelectric properties.[21] However, high-throughput experimental approach has not been attempted so far for the solar-driven thermochemical hydrogen production (STCH) oxide perovskite materials.

In this study, we screen compositional space and operating temperature of perovskite materials relevant to STCH applications using high-throughput experimental method. We synthesize thin film combinatorial libraries of $(Ba,Sr)MnO_3$, $Ba(Ce,Mn)O_3$, $(Sr,Ce)MnO_3$, and $Ba(Ce,Mn)O_3$ - $(Sr,Ce)MnO_3$ oxide alloys (solid solutions) covering a very large compositional space, and anneal the resulting films at different temperatures to explore the stability of various perovskite-related structures. From the high throughput screening of 400 different samples with $ABO_3$ stoichiometry and Ba, Sr, Ce, Mn elements, we find that the crystal structures are strongly dependent on chemical composition and annealing temperature. For example, Sr substitution on A-site of $Ba_{1-x}Sr_xMnO_3$ leads to hexagonal perovskite-like polytypic transformation structure after annealing at elevated temperature. In addition, Ce substitution on B-site of $BaCe_xMn_{1-x}O_3$ leads to the formation of a hexagonal perovskite-related polytype in a narrow composition and temperature space, whereas Ce substitution on A-site of $Sr_{1-x}Ce_xMnO_3$ leads to phase transition to tetragonal perovskite in a wide composition-temperature range. We also find that the 25% Ce substitution on both A-site and B-site of $Ba_{1-x}Sr_xMnO_3$, along the $BaCe_xMn_{1-x}O_3$ - $Sr_{1-x}Ce_xMnO_3$ pseudo-binary tie-line, further enhanced the stability of the tetragonal perovskite structure, indicating it may be a promising material for STCH applications due to broad composition-temperature range of stability.

## 2. Experimental Methods

Thin films with continuous composition spreads were grown using combinatorial Pulsed Laser Deposition (PLD). As shown in Figure 1A, to achieve composition spreads during thin film depositions, two targets were moved to the laser position in alternating fashion, accompanied by the 180° substrate rotation, resulting in a one-dimensional gradient in composition across the substrates. Single compositions were deposited from single targets by the same method. The alternating layers were grown by about 45 laser pulses in each deposition cycle and about 750 cycles for each deposited film. The composition of the of the resulting combinatorial thin film sample libraries covered approximately 50 at.%, and the thickness ranged from 350 to 600 nm as shown in Figure 1B. This thickness gradient resulted from the difference in number of laser pulses received by each side of the substrate necessary for varying the composition spread. Such thickness gradient would influence XRD peak intensity, but not the phase fractions and peak positions reported in this paper.

The combinatorial thin films of $BaMnO_3 - SrMnO_3$, $BaMnO_3 - CeO_2$, $SrMnO_3 - CeO_2$, and $BaCe_{0.25}Mn_{0.75}O_3 - Sr_{0.75}Ce_{0.25}MnO_3$, as well as single-composition films of $BaMnO_3$, $SrMnO_3$, $CeO_2$ were grown on sapphire substrates (50.8 mm in diameter, UniversityWafer, Inc.) from custom PLD targets with diameters of 25.4 mm. These custom PLD targets of $BaMnO_3$ (BMO), $SrMnO_3$ (SMO), $BaCe_{0.25}Mn_{0.75}O_3$ (BCM25), $Sr_{0.75}Ce_{0.25}MnO_3$ (SCM25), and $CeO_2$ (FL) were fabricated using a sol-gel modified Pechini method[22] by O'Hayre group at the Colorado School of Mines. The oxygen pressure in the chamber was set at 40 mTorr and the targets were ablated with a 248 nm KrF laser (Coherent Compex 205) with of ~1.8 J/cm$^2$ energy density, 30 ns pulse width, and 20 Hz repetition rate. The film growth was conducted without intentional heating of the substrate. Upon completion of the deposition, each composition gradient film library was sequentially annealed at 650 °C, 850 °C, 950 °C, 1050 °C and 1150 °C for 2 hours in box furnace in air, with heating and cooling rates of 10 °C/min.

After each annealing step, the combinatorial sample libraries with continuous composition spreads were characterized at 40 positions using spatially-resolved characterization methods to connect observed trends in crystal structure and phase distribution to changes in film composition and annealing temperature. The 40-point combinatorial grid included 4 rows with 12.5 mm spacing, as well as 9 columns (1st/4th row) and 11 columns (2nd/3rd row) with 4 mm spacing. The X-ray diffraction (XRD) used to determine the crystal structure was performed on a Bruker DISCOVER D8 diffractometer using θ-2θ geometry with Cu Kα (λ = 1.54 Å) radiation and a high resolution 2D detector. X-ray fluorescence (XRF) used to determine both the composition and thickness of the films was performed on an Fischerscope X-ray XUV 773 instrument with a Rh X-ray beam approximately 2 mm in diameter, with sensitivity factors calibrated by commercial standards (MicromatterTM). To analyze the substantial amount of data produced by the combinatorial experiments, open-source CombIgor package[23] for Igor Pro was used and the resulting data is available through High Throughput Experimental Materials (HTEM)[24] Database.

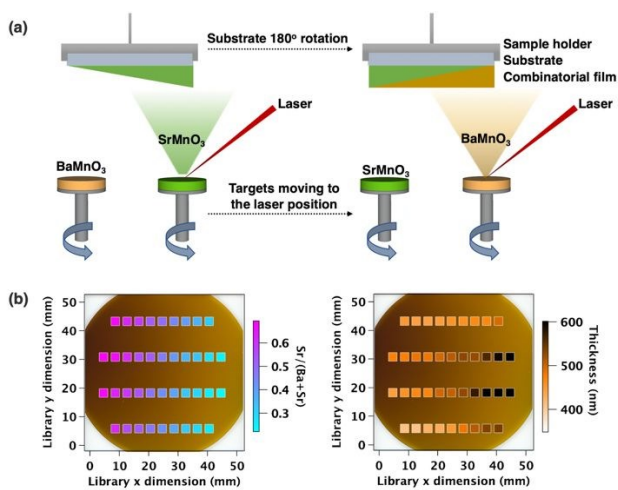

Figure 1. High-throughput experimental methods: (a) Combinatorial pulsed laser deposition geometry utilized in this study on the example of a compositionally graded Ba$_{1-x}$Sr$_x$MnO$_3$ film, and (b) a photo of a combinatorial Ba$_{1-x}$Sr$_x$MnO$_3$ film with a continuously varying composition and thickness measured at 40 positions across the 2"-diameter sapphire substrate using spatially-resolved characterization methods.

Table 1 – Summary of deposition conditions for BSM, BCM, SCM, and BSCCM thin film libraries with composition gradients. The films were deposited on sapphire substrates at room temperature and sequentially annealed in the 650-1150°C temperature range.

| Film Composition | Target materials | Structures observed |
|---|---|---|
| (Ba,Sr)MnO$_3$ | BaMnO$_3$, SrMnO$_3$ | 2H, 4H, 6H |
| Ba(Ce,Mn)O$_3$ | BaMnO$_3$, BaCe$_{0.25}$Mn$_{0.75}$O$_3$, CeO$_2$ | 2H, 10H, FL |
| (Sr,Ce)MnO$_3$ | SrMnO$_3$, Sr$_{0.75}$Ce$_{0.25}$MnO$_3$, CeO$_2$ | 4H, 3C, FL |
| 25% Ce:(Ba,Sr)MnO$_3$ | BaCe$_{0.25}$Mn$_{0.75}$O$_3$, Sr$_{0.75}$Ce$_{0.25}$MnO$_3$, | 2H, 4H, 6H, 10H, 3C, FL |

## 3. Results

The family of ABO$_3$ materials with perovskite-related structures presents a large variety of structures depending on the relative size of the cations, illustrated in Figure 2. These differences can be quantified by the Goldschmidt tolerance factor,[25] $t = \frac{r_A + r_O}{\sqrt{2}(r_B + r_O)}$ where $r_A$, $r_B$, and $r_O$ correspond to the average ionic radii of the A, B and O elements, but alternative perovskite stability metrics have been also recently developed.[26] The cubic perovskite ABO$_3$ structures, and its various tetragonal or rhombohedral distortions, occur when the Goldschmidt tolerance factor is t < 1 (Figure 2a). These cubic and distorted ABO$_3$ perovskites feature only corner-sharing BO$_6$ octahedra with cubic close-packed AO$_3$ layers in an AB arrangement. The hexagonal ABO$_3$ perovskite-like polytypes occur for larger A-site cations and/or small B-site cations, with tolerance factor is $t > 1$ (Figure 2b). These hexagonal ABO$_3$ perovskite-like polytypes often contain a mixture of face- and corner-sharing BO$_6$ octahedra along the hexagonal c-axis derived from the rotations of BO$_6$ octahedra, lowering the unit cell symmetry from cubic. The specific hexagonal polytype (e.g., 2H, 4H, 6H, 10H in Fig. 2b) depends on the corner/hexagonal ratio corresponding to cubic closed-packed (c) and hexagonal closed-packed (h) AO$_3$ layers in the perovskite structure. The summary for the different hexagonal polytypes, tetragonal perovskite, and fluorite structure is presented in Table 2.

Among the Ba-Sr-Ce-Mn-O material system considered in this study, the hexagonal perovskite-like polytypes are observed in (Ba,Sr)MnO$_3$ and in Ba(Ce,Mn)O$_3$ materials. BaMnO$_3$ has a pure hexagonal 2H polytype structure with space group of $P6_3/mmc$ which consists of infinite chains of face-sharing octahedra, resulting from the largest tolerance factor ($t$ = 1.103) due to the largest Ba ionic size.[27] The face-shared connectivity of this 2H structure changes with substitution of cations and after elevated annealing temperatures which give rise to an increase of the ratio of cubic (c) to hexagonal (h) closed-packed layer in the sequence of alternation of cation layers. The substitution of the smaller Sr cation on A-site of $BaMnO_3$ leads to the decrease of tolerance factor, driving the phase transition from pure hexagonal (2H) to the 4H which has 50% corner-shared connectivity with an (*chch*) stacking sequence[28] and further transition to 6H, which has 66% corner-sharing with a (*cchcch*) sequence,[29] after annealing at elevated temperatures. The 10H structure is found in a narrow composition-temperature stability range of $Ba(Ce,Mn)O_3$ material, and in other substituted $BaM_yMn_{1-y}O_3$ (M = Sb, In, Ln, etc.) systems[30-32] that can only be formed at the Mn/M=25/75 ratio on the B-site of the 10H perovskite structure. This 10H polytype structure consists of a (*chhhc*)$_2$ stacking sequence of the $AO_3$ layers with face-sharing octahedral formation of $[B_4O_{15}]$ tetramers that are corner-sharing with $BO_6$ octahedra along the c-axis. The Mn ions sit on the face-sharing octahedra while the large M ions occupy the corner-sharing octahedra which plays a key role in the redox and magnetic behaviors.

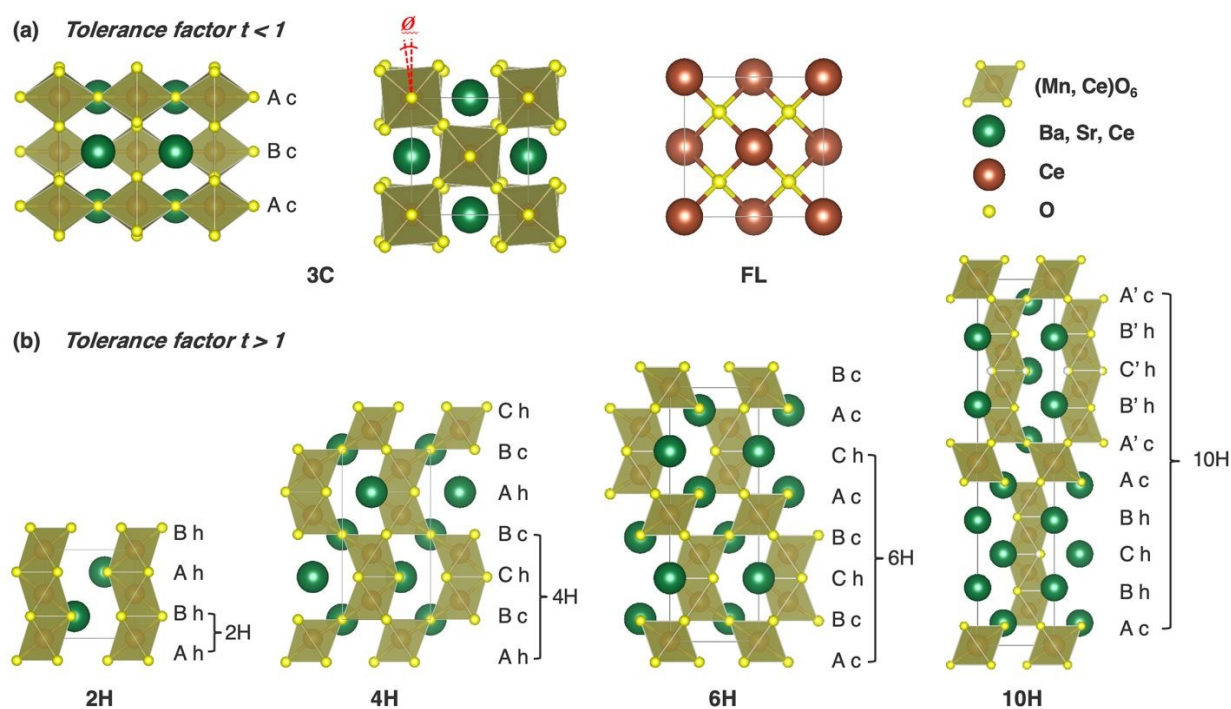

Figure 2. Crystal structures: (a) Tetragonal perovskite 3C viewed along *b* and *c* showing corner-sharing octahedra connectivity and the slight tilting of the $BO_6$ octahedra. The cubic fluorite structure (FL) for the pure $CeO_2$ crystal structure. (b) Hexagonal polytypes structures viewed along *a*-axis, showing the possible arrangements and the connectivity of the $MnO_6$ octahedra with the $AO_3$ stacking sequence. Only face-sharing octahedra leads to a 2H and a mixture of corner- and face-sharing octahedra favors a 4H, 6H, and 10H polytypes. The *c* and *h* denote cubic and hexagonal closed-packed layers, respectively.

Tetragonally distorted perovskites are observed in (Sr,Ce)MnO$_3$ oxides studies here. This tetragonal perovskite structure has space group *I*4/*mcm* (3C) with slight distortion of the $BO_6$ octahedra along the c-axis while maintaining their corner-sharing connectivity.[33] This structure is obtained by the substitution of smaller $Ce^{3+}$ ion on A-site of 4H $SrMnO_3$ material which leads to decrease in the tolerance factor. In this $Sr_{1-x}Ce_xMnO_3$ material system, the substitution of the $Sr^{2+}$ by $Ce^{3+}$ leads to mixed valence $Mn^{3+}$ and $Mn^{4+}$ to compensate the charge balance and thus, the tolerance factor is reduced to $t < 1$ by both reducing A-site cation size and increasing B-site size. In addition to these perovskite and perovskite-related structures, $CeO_2$ phase with a cubic fluorite structure (FL) of space group $Fm\overline{3}m$ is observed in higher Ce substitution level and after annealing at elevated temperatures in Ba-Sr-Ce-Mn-O material system. Figure 2a shows the structure of the stoichiometric $CeO_2$ with the 4-coordinated oxygen and the 8-coordinated cerium. Cerium is at the center of tetrahedron whose corners are occupied by oxygen atoms. The structural stability of all these phases depends on composition and annealing temperature. In the following sections we describe the stable structures with respect to its constituent elements and annealing temperature in the Ba-Sr-Ce-Mn oxide material system.

Table 2 – Crystal structure, stacking sequence, unit cell parameters for the hexagonal polytypes (2H, 4H, 6H, 10H) and the tetragonal perovskite (3C) and fluorite (FL) structures. The lattice parameters and unit cell volumes are taken from the Inorganic Crystal Structure Database (ICSD) for the example composition.

| Denotation | Crystal system | Space group | ABC | h-c notation | $a$ [Å] | $c$ [Å] | V [Å$^3$] | Example composition |
|---|---|---|---|---|---|---|---|---|
| 2H | Hexagonal | $P6_3/mmc$ | AB | h | 5.69 | 4.81 | 135 | $BaMnO_3$ |
| 4H | Hexagonal | $P6_3/mmc$ | ABCB | ch | 5.57 | 9.18 | 247 | $Ba_{0.5}Sr_{0.5}MnO_3$ |
| 6H | Hexagonal | $P6_3/mmc$ | ABCBAC | cch | 5.43 | 13.40 | 342 | $SrMnO_3$ |
| 10H | Hexagonal | $P6_3/mmc$ | ABCBA | chhhc | 5.79 | 23.90 | 693 | $BaCe_{0.25}Mn_{0.75}O_3$ |
| 3C | Tetragonal | $I4/mcm$ | AB | c | 5.42 | 7.73 | 227 | $Sr_{0.75}Ce_{0.25}MnO_3$ |
| FL | Cubic | $Fm-3m$ | ABC | c | 5.41 | | 158 | $CeO_2$ |

### 3.1 $Sr_xBa_{1-x}MnO_3$ (SBM)

Figure 3a shows a color scale map of diffracted intensity as a function of Sr/(Ba+Sr) atomic ratio and diffraction angle, to determine the effect of A-site substitution with alkaline earth metals on the compositionally graded $Ba_{1-x}Sr_xMnO_3$ (BSM) films measured after annealing at different temperatures. The XRD measurements after annealing in the entire range of temperatures shows a gradual peak shift to higher angles with increasing Sr-fraction due to the A-site substitution of larger $Ba^{2+}$ by smaller $Sr^{2+}$. This size difference of the A-site cation is believed to be largely influencing the formation of hexagonal perovskite-related structures (Figure 2). The $BaMnO_3$ crystallizes the 2H hexagonal perovskite structure containing face-sharing octahedra,[27] while $SrMnO_3$ adopts the 4H structure containing both corner- and face-sharing octahedra[28] after annealing in the range of 650 - 1050 °C . After annealing at 950°C (Figure 3c), the data clearly shows the polytypic transition from 2H to 4H polytype with increasing Sr substitution level on $Ba_{1-x}Sr_xMnO_3$. The hexagonal perovskite-related structures are formed by expanding A-site space along with shortening Mn-Mn distance which is associated with the centers of the face-sharing octahedra. The Mn-Mn distance in face-sharing polyhedra in 2H-BaMnO$_3$ is shorter (2.41 Å) than that of 4H-SrMnO$_3$ (2.50 Å).[34] The hexagonal perovskite-related polymorph stability can also be affected by annealing temperature. As shown in Figure 3a, the 2H (2H-BM, 2H-BSM) structure, which was dominant after lower annealing temperatures, gradually transforms to 4H (4H-BSM, 4H-SM) structure after elevated annealing temperature, as indicated by the new peak at 35.1° in Figure 3a.

Figure 3b displays the summary of the major peak position as a color scale and a 4H peak intensity as a marker size, clearly showing the transformation from 2H and 4H after elevated annealing temperature. Subsequently, the 4H is further converted to 6H after annealing at 1150 °C temperature. The polytypic transformation follows 2H → 4H → 6H sequence, well known in hexagonal transition metal oxide perovskite related materials at elevated temperature and pressure.[35] A likely driving force for this temperature-induced increase of a cubic/hexagonal ratio from pure hexagonal (2H) to corner/face-sharing ratio of 1:1 (4H) and 2:1 (6H), is that the corner-sharing octahedra has larger flexibility and entropy at a given temperature than face-sharing octahedra.[34] Ba-deficient Ba$_{6.3}$Mn$_{24}$O$_{48}$ phase was also observed after annealing at 1150 °C temperature (Figure 3a), which may be caused by the formation of BaAl$_2$O$_4$ due to reaction with Al$_2$O$_3$ substrates, and/or by volatile character of barium containing compounds.[36] It has been reported,[37] that the 6H transforms to 3R (3C) above 1500 °C but we have not observed this phase because it is beyond the annealing temperature region that we accessed.

The BSM film color (Figure 3d) also changes in connection with the change in the phases as a function of chemical composition and annealing temperature. While 2H-BM has light brown color and 4H-SM has dark brown color after annealing at a low temperature, the film became much darker over the whole film surface when 4H-SM appears after annealing at 950 °C. The color subsequently changed to tan in the region of 6H at 1150 ºC. The observed dark brown of 4H-SM thin film is consistent with the bulk powder color which was reported in literature.[38]

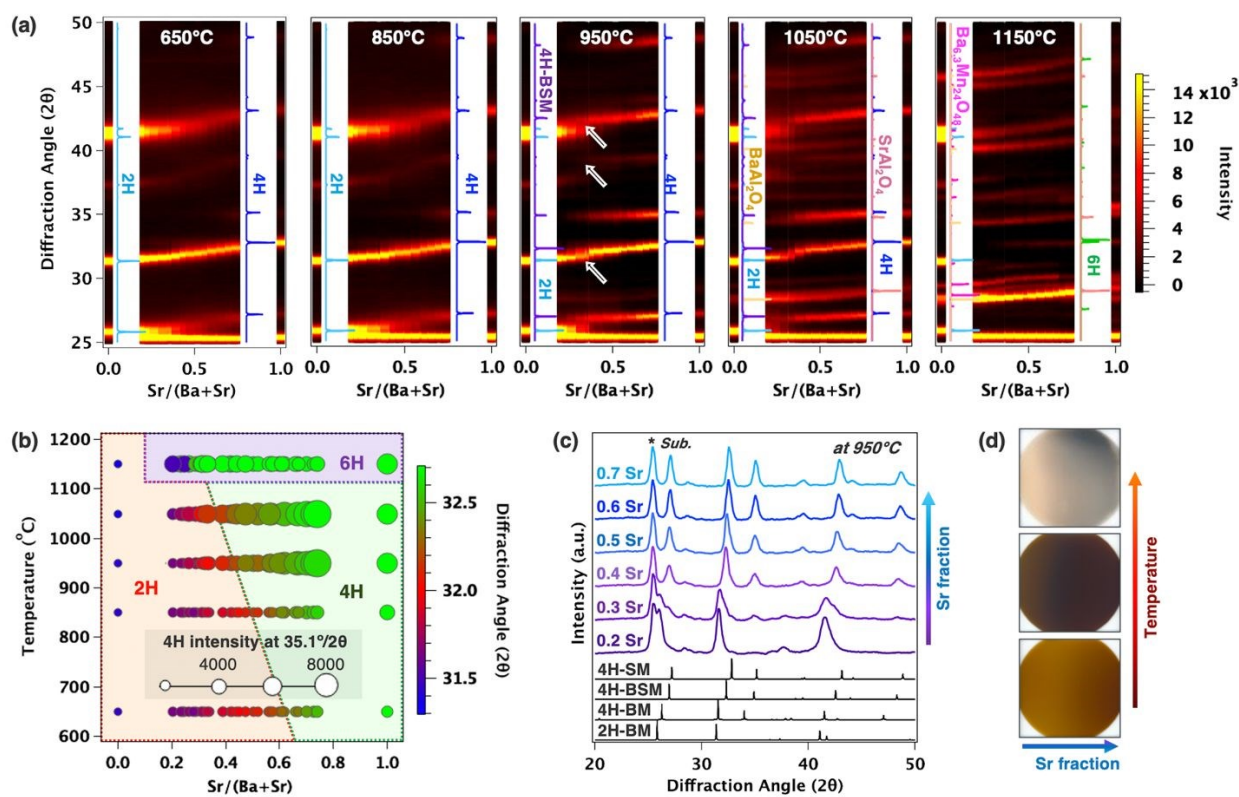

Figure 3. Ba$_{1-x}$Sr$_x$MnO$_3$ (BSM) films: (a) Color scale maps of diffracted intensity as a function of Sr/(Ba+Sr) atomic ratio vs. diffraction angle after annealing at different temperatures, with arrows indicating 2H-4H transformation. (b) summary of data for different structures as a function of chemical composition and annealing temperature. (c) XRD patterns for different Sr-fraction after annealing at 950 ºC, with * indicating sapphire substrate. (d) BSM film color measurements

### 3.2 BaCe$_x$Mn$_{1-x}$O$_3$ (BCM)

To explore effect of Ce substitution on the BaMnO$_3$, XRD profiles of BaCe$_x$Mn$_{1-x}$O$_3$ films plotted as a color scale map of diffracted intensity as a function of Ce/(Mn+Ce) ratio after annealing at different temperatures are in Figure 4a. Overall, the 2H-BM structure was observed at low Ce-fraction as indicated by XRD peak at 31.4º. However, the primary peak of 2H slightly shifted to lower angle as the Ce-fraction increased, which can be indicative of the formation of 10H-polytype, with additional peaks at 28.5º and 42.6º. Although the peak at the lower 28.5º angle coexists with FL, the 42.6 º peak at higher angle (marked by dotted line) clearly indicates the presence of the 10H structure.

All the XRD data are summarized in Figure 4b that shows the primary peak position as a function of Ce-fraction and annealing temperature. This summary plot shows an interesting region of 0.2-0.5 Ce-fraction after 850 - 1050°C annealing temperature (green shade) where the primary peak of 2H is considerably shifted to lower angle consistent with 10H polytype structure. The XRD patterns at fixed 30% Ce-amount (Figure 4c) further supports the polytypic transformation from 2H to 10H after annealing at 950°C and 1050°C. It is likely that excess Ce above 25% is accommodated by a secondary phase formation such FL even though the 10H peaks are observed up to 50% global Ce composition after annealing at 950 °C. This is because the 10H has been reported to form a line compound with an exact Ce/Mn ratio of 0.25/0.75 due to size constraints on the B-site.[39] The 10H proportion reduced after 1050 °C while the intensity of FL increases even at lower Ce-fraction, as shown in bigger marker size in Figure 4b which is indicate the increased FL intensity.

Many other perovskite-related compounds with 10H-polytype structure in the $Ba_5M_{1-x}Mn_{4+y}O_{15-\delta}$ (M = In,[30] Ln,[31] Sb[32]) material systems have been reported, as synthesized in bulk form by solid state reaction. These studies report the observation of 10H-polytype in a very narrow range of composition. For example, 10H-$Ba_5In_{1-x}Mn_{4+y}O_{15-\delta}$ structure was only observed at $x = 0.07$ and $y = 0$ in the investigated composition range of $x = 0 \sim 0.08$ and $y = -0.08 \sim 0.08$ at 1300 ºC in air. On the other hand, composition screening over the broad range in this study shows that 10H-BCM denoted by $Ba_5Ce_{1-x}Mn_{4+y}O_{15-\delta}$ structure can be observed in $x$ = -1.0 ~ 0 and $y$ = -1.0 ~ 0, in the investigated composition range of $x$ = -3.25 ~ 1 and $y$ = -3.25 ~ 1 at 850-1050 °C range in air, albeit with secondary phases such as 2H and FL. In particular, mostly phase-pure 10H structure is observed in $x$ = -0.8 ~ -0.6; $y$ = -0.8 ~ -0.6 at 950 °C in air. This comparison between $Ba_5Ce_{1-x}Mn_{4+y}O_{15-\delta}$ and $Ba_5In_{1-x}Mn_{4+y}O_{15-\delta}$ illustrates an advantage of the composition spread thin film deposition method in studying compositionally complex perovskite related materials.

The pictures of BCM films shown in Figure 4d were taken from the combinatorial $BaCe_yMn_{1-y}O_3$ samples in the region of y = 0.3-0.6 where most phases such as 2H, 10H, and FL structures coexist. At 650 °C, mainly 2H was dominant in 0.3-0.6 Ce-fraction as shown in Fig. 3a-b, and the film largely appeared pale-green which is in good agreement with a color of bulk powder.[27] After annealing at 950 °C, 10H was dominant in 0.3-0.5 Ce-fraction while the FL became stronger after 0.5 Ce-fraction, with the 10H showing bright-orange color. The BCM film changed to mostly white after annealing at 1150 °C where FL is dominant.

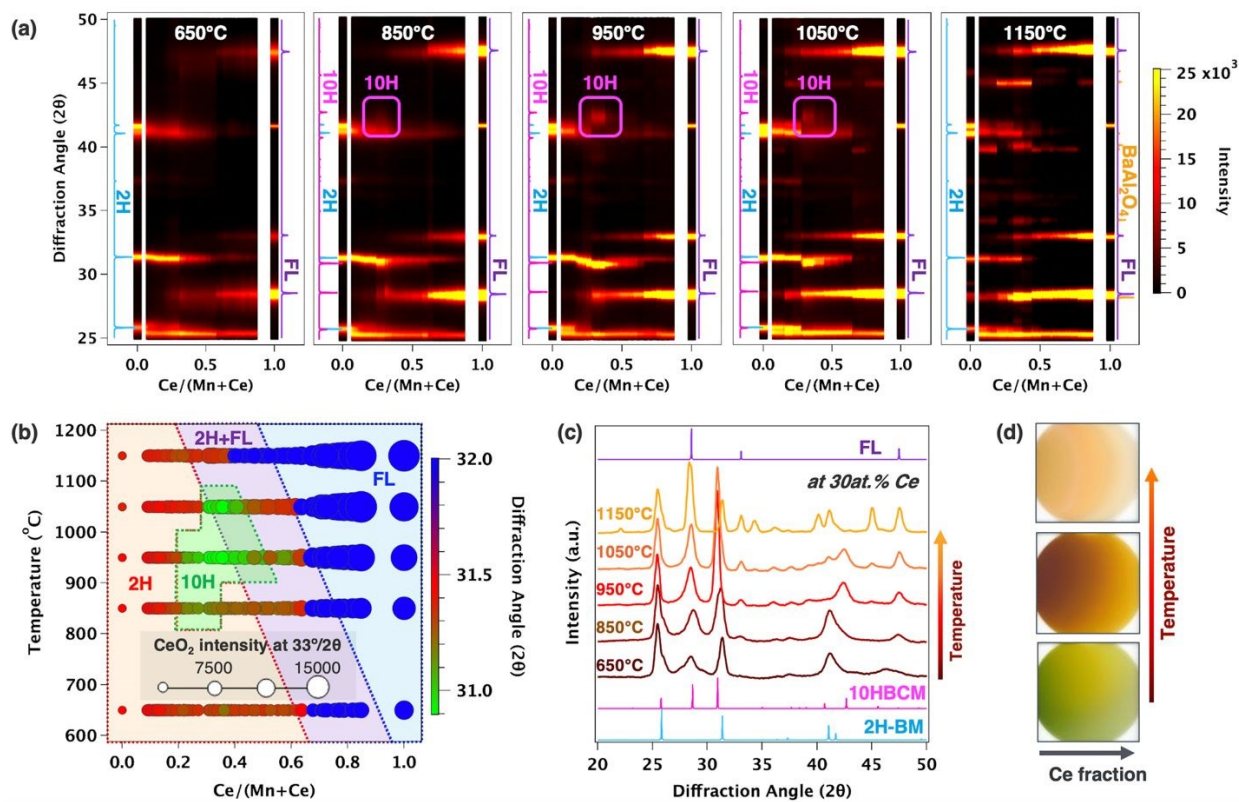

Figure 4. BaCe$_y$Mn$_{1-y}$O$_3$ (BCM) films: (a) Color scale maps of diffracted intensity as a function of Ce-fraction and diffraction angle after annealing at different temperatures, (b) summary of data for different structures as a function of chemical composition and annealing temperature, and (c) XRD patterns at fixed 30 at. % Ce after annealing at different temperatures. (d) BCM film color measurements.

### 3.3 Sr$_{1-x}$Ce$_x$MnO$_3$ (SCM)

To determine the effect of Ce substitution on A-site of SrMnO$_3$, XRD patterns were plotted as a function of Ce/(Sr+Ce) fraction after annealing at different temperatures, as shown in Figure 5a. After lower annealing temperatures (650-850 °C), XRD profiles of unsubstituted SrMnO$_3$ are consistent with the 4H-SrMnO$_3$. As the Ce-fraction increases, the peak intensity of the 4H gradually decreases while FL phase becomes stronger. The distorted 3C perovskite of interested here is not formed after this lower annealing temperature.

The perovskite 3C-SCM phase started to be observed above 950 °C. Although the primary peak of 3C structure at 32-33°/2θ coexists with those of 4H and FL, these individual peaks can be distinguished by a closer examination of the XRD profiles shown in Figure 5b. We can also identify the 3C phase by a peak at 40.7°/2θ where only the 3C peak exists, and thus the intensity of this peak is represented by marker size in Figure 5b. The distorted 3C perovskite structure appears as soon as Ce was substituted in SrMnO$_3$ after annealing at 950 °C and persists up to 0.45 Ce-fraction. However, the 3C coexists with 4H at low Ce-fraction and with FL at high Ce-fraction. This result indicates that the Ce ions are not fully substituted in the Sr-site at 950 °C. After a higher annealing temperature of 1050 °C, the 3C phase fraction becomes larger while both 4H and FL phase fractions are reduced particularly in the 0.10 – 0.25 Ce-fraction range, leading to almost phase-pure 3C structure in this composition-temperature region. This suggests that the 3C-SCM is more stable in this higher temperature range compared to the 10H-BCM. After even higher annealing temperatures of 1150 °C, the SCM decomposed to SrO$_2$, possibly due to interaction with the substrate.

Figure 5c shows that as Ce-fraction increases the 3C peak shifted to lower angles (larger lattice constant), which is unexpected if smaller Ce$^{4+}$ or Ce$^{3+}$ ions are partially substituting the larger Sr$^{2+}$ ions. This observation can be explained if the substitution with Ce$^{3+}$ for Sr$^{2+}$ creates Mn$^{3+}$ for charge compensation, leading to lattice expansion due to the larger difference in the ionic radii between Mn$^{3+}$ (0.645 Å) and Mn$^{4+}$ (0.530 Å) than between Sr$^{2+}$ (1.44 Å) and Ce$^{3+}$ (1.34 Å). According to previous publications,[40] the Ce oxidation state and site preference may vary with increasing Ce substitution level in (Sr$_{1-x}$Ce$_x$)MnO$_3$, with the Ce$^{4+}$ oxidation state decreasing on B-site at x < 0.2, and Ce$^{3+}$ ions substituting on A-site while Mn oxidation state decreasing on B-site at x > 0.2. Other published bulk studies showed that when high amount of Ce was substituted on B-site during SrCe$_y$Mn$_{1-y}$O$_3$ synthesis, this perovskite-related stoichiometry was unstable and instead a layered perovskite Ce$_x$Sr$_{2-x}$MnO$_4$ formed.[13]

Our results presented in Figure 5 are largely consistent with published literature where the Ce substitution on A-site of 4H-SrMnO$_3$ has been reported to form a tetragonal symmetry Sr$_{1-x}$Ce$_x$MnO$_3$ in a wide range of Ce concentrations (0 ≤ x ≤ 0.5) in bulk form by solid-state reaction methods.[41-43] These reports showed that with increasing Ce-contents and annealing temperature, the tetragonal perovskite SCM is stabilized as accompanied by appearance of secondary phase of CeO$_2$, which is similar trend to our results from the thin film combinatorial PLD method reported here. However, while the bulk powder 3C-SCM phase is usually obtained by calcination at high temperature over 1400 °C for a long time (over 10 h) by solid-state reaction method, PLD can obtain the 3C-SCM phase in thin film after annealing at much lower temperature in 950-1150 °C range.

The pictures of Sr$_{1-x}$Ce$_x$MnO$_3$ films are shown in Figure 5d, in the region of x = 0.2-0.6 where all the phases such as 4H, 3C, and FL exist. After annealing at 650 °C, 4H and FL phase color matches well with the results shown in Figure 4d and discussed above. At 950 °C the 3C phase has dark-brown color, followed by black color at 1150°C, as expected for the phase transition from hexagonal to tetragonal phases leading to the color change.

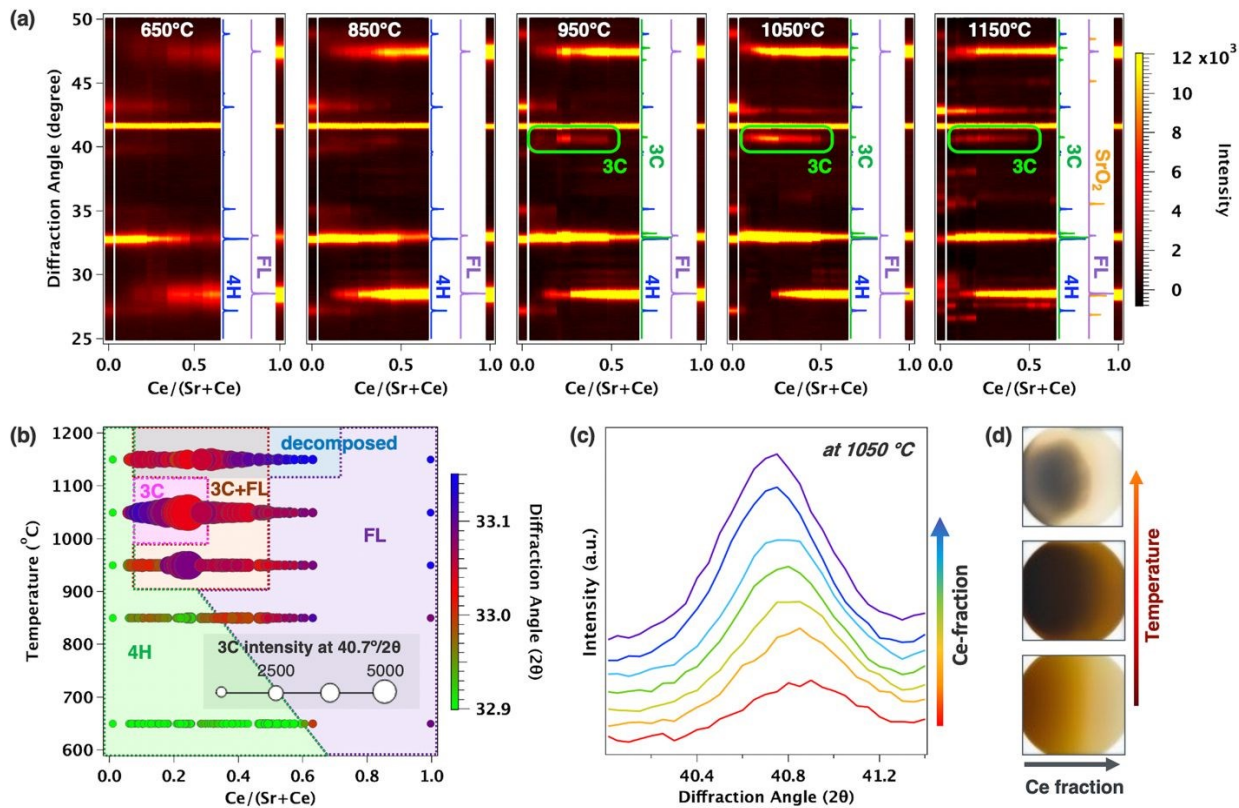

Figure 5. Sr$_{1-x}$Ce$_x$MnO$_3$ (SCM) films: (a) Color scale maps of diffracted intensity as a function of Ce/(Sr+Ce) atomic ratio and diffraction angle at different annealing temperatures, (b) summary of data for different structures as a function of chemical composition and annealing temperature, (c) close examination of XRD patterns after annealing at 1050 °C that shows the 3C-SCM peak shift. (d) SCM film color measurements.

### 3.4 (Ba,Sr)MnO$_3$ with 25% Ce substitution (BSCM)

In order to determine the effects of both A- and B-site substitution on the (A′$_x$A″$_{1-x}$)(B′$_x$B″$_{1-x}$)O$_{3-\delta}$, 25% Ce-substituted Ba$_{1-x}$Sr$_x$MnO$_3$ combinatorial film (BSCM) was grown by PLD using BCM25 and SCM25 targets. As the both targets contain 25% Ce-content, we plot the composition XRD profiles as a function of Sr/(Ba+Sr) atomic ratio at fixed 25% Ce-content after annealing at different temperatures, in a pseudo-binary oxide alloy fashion in Figure 6a.

Figure 6a shows that overall the primary peaks of the 2H-BM structure gradually shifted to higher angles to be 4H-SM structure as increasing Sr fraction in Ce:Ba$_{1-x}$Sr$_x$MnO$_3$, and secondary phases such 10H, 3C, and FL appeared depending on the annealing temperature. At 650 °C, the 2H peaks gradually shifted to higher angle with increasing Sr-fraction and reached to the 4H peak positions at 1.0 Sr-fraction with a minor FL secondary phase. This Ce:Ba$_{1-x}$Sr$_x$MnO$_3$ result (Fig. 6) is consistent with the results of BaCe$_y$Mn$_{1-y}$O$_3$ (Fig. 4) and Sr$_{1-x}$Ce$_x$MnO$_3$ (Fig. 5) at $x = y = 0.25$. In the annealing temperature range of 850-1050 °C, the 10H structure started to appear from 0% up to 25% Sr-content, as shown in the peak shift in the range of 30-32°/2θ and a new peak at 42°/2θ (Figure 6b). This Ce:Ba$_{1-x}$Sr$_x$MnO$_3$ result shows promise for substituting foreign cations (e.g. up to 25% Sr) on A-site of 10H-BaCe$_{0.25}$Mn$_{0.75}$O$_3$ to increase its narrow composition stability range.

The tetragonal 3C-Sr$_{0.75}$Ce$_{0.25}$MnO$_3$ can accept even larger amount of Ba-substitution on A-site, as shown in Figure 6b. The 3C structure is observed up to 23% → 35% → 64% Ba-substitution on A-site of Sr$_{0.75}$Ce$_{0.25}$MnO$_3$ after the annealing temperature is increased from 950 °C to 1050 °C to 1150 °C. Interestingly, if the 4H-polytype exists the distorted 3C phase fraction is significantly reduced, whereas if the 3C phase becomes stronger the 4H-polytype almost disappears. Furthermore, FL intensity was also reduced where 3C intensity was strong, however the FL phase becomes stronger at elevated annealing temperature. Figure 6c shows the phase transformation following 10H → 2H → 4H → 3C sequence with increasing Sr-fraction in 25% Ce-substituted Ba$_{1-x}$Sr$_x$MnO$_3$ after annealing at 950°C.

Figure 6b summarized these phase transitions on the color scale by diffraction angle in the range of 31.2 – 32.7°, where all the primary peaks of 2H, 4H, 10H, and 3C coexist, and where we can determine the dominant phase by the diffraction angle value. In addition, the marker size in Figure 6b further identifies 10H and 3C structures by plotting as a marker size their individual peaks intensity at 40.7° and 42.6 ° respectively. These Ce:Ba$_{1-x}$Sr$_x$MnO$_3$ results clearly indicated that the 10H structure is stable in a narrower composition range at lower annealing temperatures, compared to the distorted 3C structure that is stable in a wider composition range and at higher annealing temperature. Compared to the Sr$_{1-x}$Ce$_x$MnO$_3$ system shown in Fig. 5, the 3C structure stability is enhanced in Ba- substituted Sr$_{1-x}$Ce$_x$MnO$_3$ system.

The Ce:Ba$_{1-x}$Sr$_x$MnO$_3$ film color of BSCM25 as shown in Figure 6d indicates that the FL color (white) is not largely observed even after annealing at 1150°C. These pictures indicate that the 25% Ce substitution leads to mostly the formation of 10H and 3C structures in Ba$_{1-x}$Sr$_x$MnO$_3$.

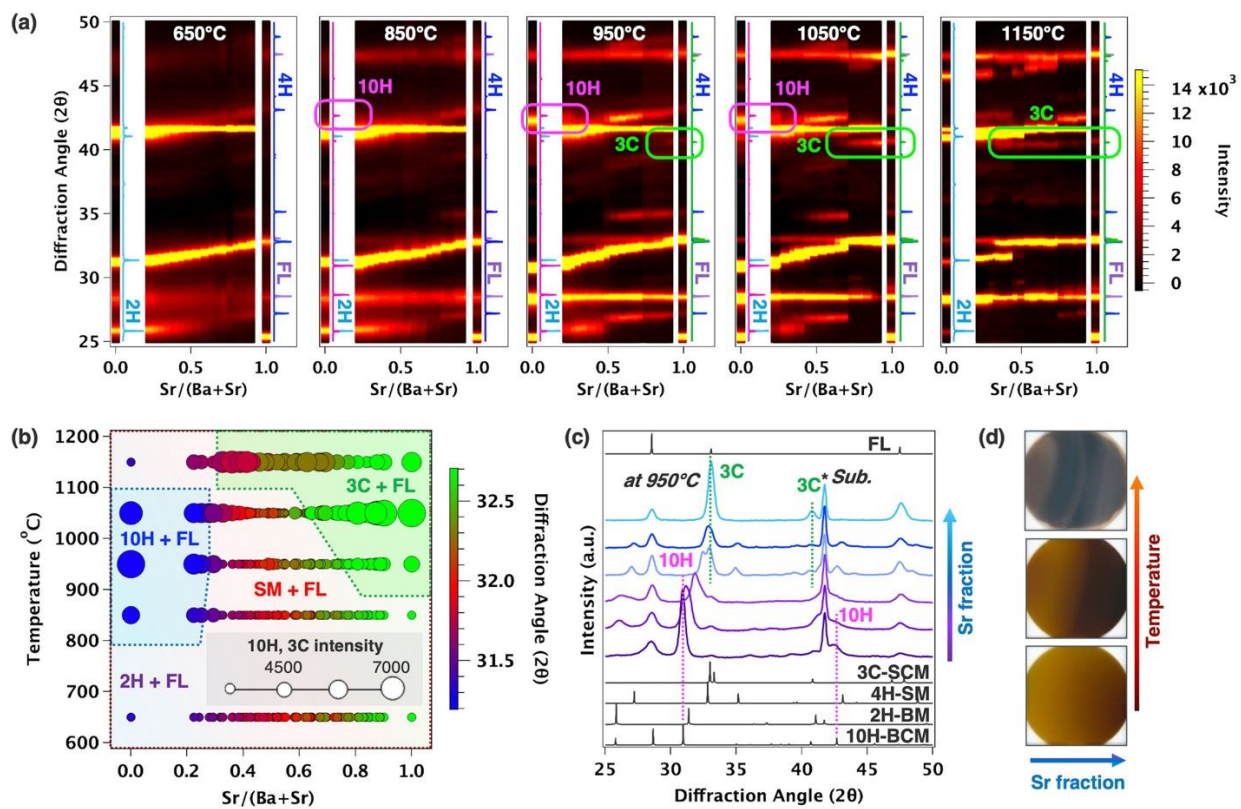

Figure 6. 25% Ce-substituted Ba$_{1-x}$Sr$_x$MnO$_3$ (BSCM25) films: (a) XRD color scale maps as a function of Sr/(Ba+Sr) atomic ratio at fixed 25% Ce-content and at different annealing temperatures, (b) summary of data for different structures as a function of chemical composition and annealing temperature. (c) XRD patterns of BSCM25 as a function of Sr content after annealing at 950°C indicating the phase change from 10H to 3C. (d) BSCM25 films color measurements.

## 4. Discussion

Temperature-composition map summarizing the results of our combinatorial work on the Ba-Sr-Mn-Ce oxides is shown in a 3-dimensional (3D) plot in Figure 7a. The end-point compounds like $BaMnO_3$ and $SrMnO_3$ as well as $CeO_2$ are located at the corners of the triangle in x-y plane, with the annealing temperature increasing along the z-axis. The single-site substituted perovskite oxides such as $BaCe_yMn_{1-y}O_3$ (BCM), $Sr_{1-x}Ce_xMnO_3$ (SCM), and $Ba_{1-x}Sr_xMnO_3$ (BSM) are located on the sides of the triangles. A more complex 25% Ce-substituted $Ba_{1-x}Sr_xMnO_3$ pseudo-binary oxide alloy (BSCM25) between BCM25 and SCM25 is shown by a line inside of the triangle parallel to the $Ba_{1-x}Sr_xMnO_3$ (BSM) line.

In this Figure 7a, the rainbow color gradient of the markers displayed represents each of these structures, proportional to a characteristic XRD peak angle displayed in the color legend, as shown on the example of 2D BM-SM projection in Fig. 7b. The yellow markers in Fig. 7b indicate 2H structure of BM composition, the color changes to green as Sr-fraction increases in the BM-SM oxides, and reaches to sky-blue indicative of 4H structure of SM composition. In addition, this 2D projection (Fig.7b) of the 3D map (Fig.7a) shows the annealing temperature dependence of these structures, with $BaMnO_3$ having a 2H structure after annealing at all the temperatures, but $SrMnO_3$ having a 4H structure after annealing at 950 – 1050 ºC and a 6H structure at 1150 ºC.

The 2H, 4H, 6H hexagonal polytypes highlighted in Figure 7b are also present in Ce-substituted $BaMnO_3$, $SrMnO_3$, and $(Ba,Sr)MnO_3$ in Figure 7a, but their proportion is reduced at higher Ce concentration and after higher annealing temperatures. Depending on the annealing temperature and Ce substitution level in the $BaMnO_3$, $SrMnO_3$, and $(Ba,Sr)MnO_3$, these 2H, 4H, 6H hexagonal polytypes are transformed to other phases such as 10H, 3C, and FL structures. For example, along the BM-Ce tie line, the structure changes with increasing Ce content from 2H to 10H within the 20-50% Ce region, and this 10H region extends into the center of the phase diagram (~25% Sr concentration). Although the 10H polytype is observed in wide range of Ce concentration after annealing at 850 ºC - 1050 ºC, secondary phases such 2H and FL structures are also observed (see Fig. 4) because of the line compound character of the 10H structure with an exact Ce/Mn ratio.

The results presented in Figure 7a also show a wide range of Ce concentration in $Sr_{1-x}Ce_xMnO_3$ with 3C structure. This tetragonal 3C perovskite is observed on the SM-$CeO_2$ tie line in Figure 7a for Ce concentration from 10% to 45% after annealing within 950 ºC - 1150 ºC temperature range, with almost pure 3C tetragonal structure between 10 and 25% Ce concentration at 950 ºC (see also Figure 5). In addition, the region in the center of the 3D phase diagram in Figure 7a shows that the 3C structure can accommodate up to 64% Ba on A-site of $Sr_{0.75}Ce_{0.25}MnO_3$. Thus, this 3C structure, with double site substitution and broad composition-temperature range of stability, may be a new candidate material for STCH applications. It is promising due to its ability to accommodate higher Ce content compared to the line compound 10H-BCM. These results show that the Ba substitution into $Sr_{0.75}Ce_{0.25}MnO_3$ permits additional Ce to be accommodated in the quaternary BSCM composition space.

It has been shown in literature that other double-site substituted $(A'_xA''_{1-x})(B'_yB''_{1-y})O_{3-\delta}$ perovskites can be promising redox materials. For example, Mn-based perovskite oxides such $La_{1-x}Ca_xMn_{1-y}Al_yO_{3-\delta}$ [44,45] and $La_{1-x}Sr_xMn_{1-y}Al_yO_{3-\delta}$ [46,47] provide high redox activity for the thermochemical water splitting cycle due to valence state variation of $Mn^{2+}/Mn^{3+}/Mn^{4+}$ on B-site caused by heterovalent substitution on A-site (i.e. Ca and Sr for La). Similarly, the oxygen reduction in $Ba_{1-x}Sr_xCo_yFe_{1-y}O_{3-\delta}$ increased with Fe- and Sr-content, leading to a greater amount of $M^{4+}$ species in the compounds, thereby favoring higher reduction yield.[48] Thus, it would be important to determine the perovskite phase stability within the double-site substituted quaternary Ba-Sr-Ce-Mn-O material system,[49] outside of the single-site substituted perovskite oxides investigated in this study. The perovskite oxides 3D map displayed in Fig. 7a can provide a starting point for accelerated development of double-site substituted BSCM perovskite material in the future studies.

STCH performance of single-site substituted perovskite oxides discussed in this paper, such as 3C-SCM and 10H-BCM, have been reported to be better than their parent compounds $BaMnO_3$, $SrMnO_3$, and $CeO_2$.[12,13] While the $CeO_2$ has a too low extent of reduction that requires very high operating temperature (> 1500ºC),[50] undoped single perovskites ($SrMnO_3$, $BaMnO_3$) show poor reaction reversibility due to exceptionally large extents of reduction.[51] In comparison, single-site substituted $BaCe_{0.25}Mn_{0.75}O_3$ perovskite line compound yields nearly 3 times more hydrogen than $CeO_2$ at reduced operating temperature (1350 ºC), but with a polymorph phase transition (12R → 10H) during thermal reduction.[12] Therefore, better structural stability of perovskite materials in a wider range of chemical composition and operating temperature is important for STCH applications. The wide range of Ce compositions in $Sr_{1-x}Ce_xMnO_3$ with 3C structure and its ability to accommodate large Ba substitution after annealing in a wide range of temperatures, both demonstrated in this paper, suggest that these single-site substituted $Sr_{1-x}Ce_xMnO_3$ oxides and related double-site substituted $Ba_{1-x}Sr_xCe_yMn_{1-y}O_{3-\delta}$ may be promising candidate materials for STCH applications.

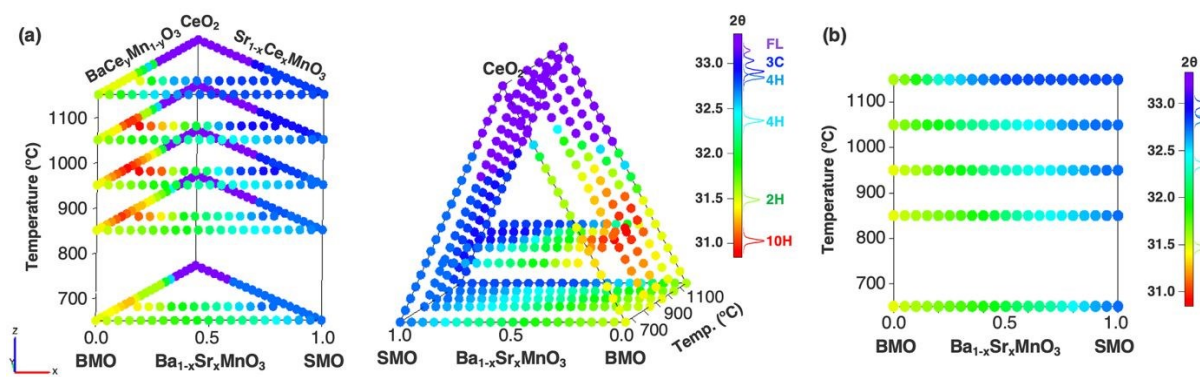

Figure 7. The Ba-Sr-Mn-Ce-O perovskite oxides. (a) Summary 3D plots of the data collected via combinatorial experiments in the 650 – 1150 ºC annealing temperature range, viewed from two different angles. The 10H-polytype (red) appears in a smaller region than 3C perovskite (blue), while FL (purple) is increasing with increasing annealing temperature. (b) Example 2D projection of the polytypic transformation sequence along the $Ba_{1-x}Sr_xMnO_3$ (BSM) oxides line following 2H → 4H → 6H transformation with increasing Sr content and increasing annealing temperature.

## 5. Summary and Conclusions

We synthesized compositionally graded $(Ba,Sr)MnO_3$, $Ba(Ce,Mn)O_3$, $(Sr,Ce)MnO_3$, and 25% Ce-substituted $(Ba,Sr)MnO_3$ thin films at ambient temperature by combinatorial PLD to investigate this family of Ba-Sr-Mn-Ce perovskite oxides with $ABO_3$ stoichiometry. The films were sequentially annealed in the temperature range of 650 – 1150 ºC, to explore the stability of various perovskite-related structures depending on site-substitution, composition, and annealing temperature by using spatially-resolved XRF and XRD measurements. Our high-throughput experimental screening study confirmed the formation of 2H, 4H, 6H, and 10H hexagonal perovskite-like polytypes in the $(Ba,Sr)MnO_3$ and $Ba(Ce,Mn)O_3$ oxides, and 3C tetragonal perovskite structure in the $(Sr,Ce)MnO_3$ oxides. The structural stability of the 3C perovskite in the composition-temperature space was considerably enhanced by Ba-substitution in the $(Sr,Ce)MnO_3$ leading to $(Ba_xSr_{1-x})_{0.75}Ce_{0.25}MnO_3$ pseudo-binary oxide alloy system. On the other hand the 10H-polytype formed by Sr substitution in the $Ba(Ce,Mn)O_3$ was confirmed in a much narrower composition-temperature stability range of this $(Ba_xSr_{1-x})_{0.75}Ce_{0.25}MnO_3$ oxides. These results indicate the promise of the 3C polytype of $(Sr,Ce)MnO_3$ and related $(Ba_xSr_{1-x})_{0.75}Ce_{0.25}MnO_3$ perovskite structures to increase the stability range of perovskite oxides for STCH applications. As a whole, the findings of the structural stability of the perovskite oxides with Ba, Sr, Mn, Ce elements and $ABO_3$ stoichiometry as a function of annealing temperature presented in this study, offer a guide for high-throughput experimental screening for the future's high-performance and stable redox materials in a broad range of energy applications.

**Conflicts of interest**

There are no conflicts to declare

**Acknowledgements**

This work was authored by the National Renewable Energy Laboratory (NREL), operated by Alliance for Sustainable Energy LLC, for the U.S. Department of Energy (DOE) under contract no. DE-AC36-08GO28308. Funding provided by the Office of Energy Efficiency and Renewable Energy (EERE) Hydrogen and Fuel Cell Technologies Office (HFTO), as a part of HydroGEN Energy Materials Network (EMN) consortium. We would like to thank Prof. Ryan O'Hayre and Dr. Michael Sanders for scientific and technique discussions and fabrication of PLD targets related to this work. The views expressed in the article do not necessarily represent the views of the DOE or the U.S. Government.